**Title Page**

**Full Title**

Automatic classification between COVID-19 pneumonia, non-COVID-19 pneumonia, and the healthy on chest X-ray image: combination of data augmentation methods

**Authors and Affiliations**

Mizuho Nishio, MD, PhD[1], Shunjiro Noguchi, MD[2], Hidetoshi Matsuo, MD[1], Takamichi Murakami, MD, PhD[1]

1 Department of Radiology, Kobe University Graduate School of Medicine, 7-5-2 Kusunoki-cho, Chuo-ku, Kobe 650-0017 JAPAN

2 Department of Diagnostic Imaging and Nuclear Medicine, Kyoto University Graduate School of Medicine, 54 Shogoin Kawaharacho, Sakyo-ku, Kyoto, 606-8507, Japan

*Corresponding author: Mizuho Nishio, MD, PhD

Department of Radiology, Kobe University Graduate School of Medicine, 7-5-2 Kusunoki-cho, Chuo-ku, Kobe 650-0017 JAPAN

Tel.: +81-78-382-6104. Fax: +81-78-382-6129

e-mail: nishiomizuho@gmail.com

**Abstract**

**Purpose**: This study aimed to develop and validate computer-aided diagnosis (CXDx) system for classification between COVID-19 pneumonia, non-COVID-19 pneumonia, and the healthy on chest X-ray (CXR) images.

**Materials and Methods**: From two public datasets, 1248 CXR images were obtained, which included 215, 533, and 500 CXR images of COVID-19 pneumonia patients, non-COVID-19 pneumonia patients, and the healthy samples. The proposed CADx system utilized VGG16 as a pre-trained model and combination of conventional method and mixup as data augmentation methods. Other types of pre-trained models were compared with the VGG16-based model. Single type or no data augmentation methods were also evaluated. Splitting of training/validation/test sets was used when building and evaluating the CADx system. Three-category accuracy was evaluated for test set with 125 CXR images.

**Results:** The three-category accuracy of the CAD system was 83.6% between COVID-19 pneumonia, non-COVID-19 pneumonia, and the healthy. Sensitivity for COVID-19 pneumonia was more than 90%. The combination of conventional method and mixup was more useful than single type or no data augmentation method.

**Conclusion:** This study was able to create an accurate CADx system for the 3-category classification. Source code of our CADx system is available as open source for COVID-19 research.

**Introduction**

The outbreak of the novel coronavirus disease (COVID-19) started in Wuhan, Hubei province, China at the end of 2019 [1], and COVID-19 spread across the world in 2020. COVID-19 is caused by a strain of coronavirus called Severe Acute Respiratory Syndrome Coronavirus 2 (SARS-CoV-2) [2]. The World Health Organization declared COVID-19 as a pandemic on March 11, 2020 [3]. COVID-19 can be detected with the use of real-time polymerase chain reaction (RT-PCR) test of SARS-CoV-2. Although the specificity of RT-PCR was sufficiently high for COVID-19, its sensitivity was relatively low in detecting COVID-19 [4]. Chest computed tomography (CT) was useful for detecting abnormal findings of COVID-19 pneumonia. It was characterized by ground-glass opacity distributed predominantly on lung peripherals [4,5]. The CT findings of COVID-19 could be regarded as distinct from viral and bacterial pneumonia.

Although the usefulness of CT for detecting COVID-19 pneumonia was shown in several studies, CT is not suitable for COVID-19 screening due to its cost and radiation exposure [6]. On the other hand, chest X-Ray imaging (CXR) is cost-effective and commonly used for screening purposes. CXR findings of COVID-19 pneumonia is characterized by the following; consolidation was the most common finding, followed by ground glass opacity; distribution of CXR abnormalities could be peripheral and lower zone distribution with bilateral involvement; pleural effusion was uncommon [7]. Compared with chest CT, the sensitivity of CXR is generally low for pulmonary diseases. Therefore, accurate diagnosis of COVID-19 pneumonia can be more challenging on CXR than on chest CT.

Computer-aided diagnosis (CADx) is being used for detection and diagnosis in several medical fields. CADx utilizes artificial intelligence methods for improving its diagnostic accuracy and robustness. Recent advances in machine learning, particularly deep learning with convolutional neural network (CNN), have shown promising performance of CADx in classifying disease patterns on medical images, such as CXR and chest CT [8–11].

The purpose of this study was to develop CADx system for classification between COVID-19 pneumonia, non-COVID-19 pneumonia, and the healthy using CXR images and CNN. Since the number of publicly available CXR images of COVID-19 pneumonia was limited, we developed the CNN model which could be accurate and robust even if the training data of CNN was small. The proposed method included the transfer learning, in which CNN models pre-trained on a large dataset is used for the improvement of accuracy and robustness [9,12]. Although this study mainly utilized a commonly used pre-trained model (VGG16), the latest CNN model (EfficientNet) was also used for transfer learning. Next, the combination of data augmentation methods was used for improving model's robustness. In addition to conventional data augmentation method (such as flipping, shifting, rotating, and etc.), mixup, and Random Image Cropping and Patching (RICAP) were used in this study [13–15]. Finally, the model was examined to evaluate whether it distinguishes COVID-19 pneumonia from both non-COVID-19 pneumonia and the healthy on CXR images.

**Material and Methods**

Our study used anonymized data collected from public datasets. Therefore, institutional review board approval was waived according to the regulations of our country. No informed consent was required.

**Dataset**

Two datasets were used: (I) one dataset for CXR images of COVID-19 and non-COVID-19 pneumonia and (II) the other for CXR images of the healthy and non-COVID-19 pneumonia. (I) The COVID-19 image data collection repository on GitHub is a growing collection of CXR and CT images of COVID-19 pneumonia [16]. In addition to

COVID-19 pneumonia, this repository contains a small number of CXR and CT images of non-COVID-19 pneumonia. (II) The RSNA Pneumonia Detection Challenge dataset available on Kaggle contains CXR images of non-COVID-19 pneumonia and the healthy [17]. Figure 1 shows representative CXR images of COVID-19, non-COVID-19 pneumonia, and the healthy.

From the dataset (I), CXR images of lateral view and CT images were excluded, and CXR images of both posterior-anterior and anterior-posterior views were included. Based on these criteria, only 215 and 33 CXR images were included from the dataset (I) for COVID-19 and non-COVID-19 pneumonia, respectively. In addition, 500 and 500 CXR images were randomly selected from the dataset (II) for the healthy and non-COVID-19 pneumonia, respectively. In order to avoid strong class imbalance, the 1000 CXR images were selected from the dataset (II). In total, 215, 533, and 500 CXR images of COVID-19 pneumonia, non-COVID-19 pneumonia, and the healthy were used for development and validation of the proposed method.

From the two datasets, patient's age and sex were collected. Table 1 summarizes the patients' characteristics and CXR attributes. The 1248 CXR images were divided into 998, 125, and 125 images for training, validation, test sets, respectively. For image normalization, CXR images were divided by 255, and pixel values of them ranged from

0 to 1.

**Deep learning model and data augmentation**

VGG16 [18] was mainly used as deep learning model for the proposed method, and transfer learning was performed for the classification of CXR images of COVID-19, non-COVID-19 pneumonia, and the healthy. To search for optimal hyperparameters of the VGG16-based model and combination of data augmentation methods, random search was performed [19]. The outline of deep learning model is shown in Figure 2.

Publicly available weights of VGG16 obtained by pre-training on ImageNet dataset were used for transfer learning. Transfer learning was performed by freezing the trainable parameters of 10 layers in VGG16. After the convolution layers of VGG16, the global averaging pooling layer, fully-connected layer, and dropout layer were added to VGG16. For the 3-category classification, the final 3-unit fully-connected layer was added after the dropout layer. Hyperparameters obtained by the random search of the VGG16-based model were as follows. The probability of the dropout layer was 0.1, and the number of units in the first fully-connected layer was 416. The input image size of VGG16 was changed to 220 × 220 pixels. RMSprop with learning rate of $1.0 \times 10^{-4}$ was used as the optimizer. The network was trained using a batch size of 8, and the number of training epochs was set to 100. The loss function was softmax loss. Early stopping was enabled using validation loss, and the patience of early stopping was set to 7.

To prevent overfitting the model training, optimal combination of the three types of data augmentation methods (conventional method, mixup, and RICAP) was also examined by the random search, and combination of conventional method and mixup were used in the proposed method of the VGG16-based model. The conventional data augmentation method included ±15° rotation, ±15% x-axis shift, ±15% y-axis shift, horizontal flipping, and 85%–115% scaling and shear transformation. The parameters of mixup was set to 0.1 [13].

The training of the model was performed using a PC with a discrete GPU (Nvidia RTX 2080 Ti, RAM 11 GB). Python (version 3.7, http://www.python.org/) was used as the programing language, and Keras (version 2.2.4, http://keras.io/) and TensorFlow (version 1.13.1, http://tensorflow.org/) were used as deep learning frameworks.

**Comparison with other pre-trained models and ablation study**

To compare with the VGG16-based model, the following four pre-trained models were used for the transfer learning: Restnet-50 [20], MobileNet [21], DenseNet-121 [22], and EfficientNet [23]. In the transfer learning using these four pre-trained models, the trainable parameters were not frozen; it was found that freezing trainable parameters in these models degraded the model performance. For the four pre-trained models, random search was also done for optimal hyperparameters and combination of data augmentation methods. For EfficientNet, the best model was selected from B0–B7 by the random search.

To evaluate the effectiveness of data augmentation methods and the freezing of trainable parameters in the VGG16-based model, the following modified models were also evaluated for VGG16-based model: (i) no data augmentation with the freezing, (ii) the conventional method only with the freezing, (iii) mixup only with the freezing, and (iv) conventional method and mixup without the freezing.

**Performance evaluation**

For each model, performance evaluation was done using the 3-category classification (ternary classification) accuracy of the test set with 125 CXR images. To assure robustness of the models, the 3-category accuracy was calculated 5 times by changing random seed, training the models, and evaluating the test set. In addition, sensitivity of COVID-19 pneumonia was also calculated using the VGG16-based model.

**Results**

Table 2 shows the results of 3-category classification between COVID-19 pneumonia, non-COVID-19 pneumonia, and the healthy for the five pre-trained models including the proposed method. The results were obtained by the random search to find the optimal hyperparameters and combination of data augmentation methods. As shown in Table 2, the mean accuracy of the VGG16-based model of proposed method was 83.7%. The mean accuracies of Restnet-50, MobileNet, DenseNet-121, and EfficientNet were lower than the VGG16-based model of propose method; the mean accuracies of these four models were less than 80%. The mean sensitivity of COVID-19 pneumonia was 90.9% for the VGG16-based model. Table 3 shows representative confusion matrix of 3-category classification in the test set.

In Table 4, the effectiveness of the data augmentation methods and the freezing of trainable parameters were evaluated in the VGG16-based model. Table 4 shows that the layer freezing was effective. The combination of two types of data augmentation methods in the proposed method was more effective than single type or no data augmentation methods.

Table S1 of Supplementary material shows the effect of RICAP obtained by the random search. Although the combination of conventional method, mixup, and RICAP was also evaluated, the combination of three methods was not as good as that of proposed method based on the results of random search. The combination of conventional method and RICAP was slightly inferior to the combination of conventional method and mixup. Therefore, the combination of conventional method and RICAP was not examined intensively in the current study.

**Discussion**

The results of this study indicate that it was possible to construct an accurate CNN model by using both the transfer learning with VGG16 and the combination of data augmentation methods. Our results show that diagnostic accuracy of the 3-category classification between COVID-19 pneumonia, non-COVID-19 pneumonia, and the healthy was more than 80% in the proposed method. In addition, the sensitivity of COVID-19 was more than 90%.

Table 4 shows that the combination of two types of data augmentation methods was more effective than single type or no data augmentation methods. Our results were consistent with the results of previous study done for bone segmentation with CNN [15]. Because the dataset of the current study was relatively small-sized (number of CXR images was 1248), it was necessary to improve the robustness of CNN models. For this purpose, the current study used the combination of data augmentation methods. The combination of conventional method and mixup was most effective.

Among the several types of pre-trained models, VGG16 was the most accurate for the 3-category classification. Although the classification accuracy of ImageNet dataset was higher in other models than that in VGG16, our results were not compatible with the results of ImageNet dataset. Since the other models are more complicated (e.g., residual learning) and/or have a large number of trainable parameters, overfitting may have occurred in the current study with the small-sized dataset. The effectiveness of pre-trained models in a small-sized dataset should be further investigated.

The layer freezing of the trainable parameters was effective only in VGG16. Network architecture of VGG16 was simpler than the other models. For example, skip connection for residual learning in the other networks may hinder the layer freezing. This may affect the usefulness of layer freezing in CNN models.

According to an article of towardsdatascience.com [24], several previous studies constructed datasets by adding pediatric CXR images of non-COVID-19 pneumonia to adult CXR images of COVID-19 pneumonia. However, when a CNN model is trained by these datasets, the model may try to distinguish between non-COVID-19 pneumonia and

COVID-19 pneumonia by checking age differences between children and adults rather than disease differentiation. Therefore, this study only included adult CXR images of the healthy and non-COVID-19 pneumonia from the RSNA dataset.

There are some limitations in our study. First, we developed and validated the proposed method using the public datasets. Overfitting may have occurred in external validation. Second, our CADx system was not used by clinicians. Clinical usefulness of our CADx system was not validated.

**Conclusion**

In conclusion, it is possible to build an accurate CADx system for 3-category classification of COVID-19 pneumonia, non-COVID-19 pneumonia, and the healthy using the proposed method. The combination of two types of data augmentation methods was more useful than single type or no data augmentation methods. We will investigate performance of our CADx system when clinical CXR images with COVID-19 pneumonia, non-COVID-19 pneumonia, and the healthy were fed to the system.

**Online content**

Source code and dataset of the current study are available at https://github.com/jurader/covid19_xp.

**Funding**

The present study was supported by JSPS KAKENHI (grant number 19H03599 and JP19K17232).

**Conflicts of Interest**

The authors declare no conflict of interest. The funders had no role in the design of the study; in the collection, analyses, or interpretation of data; in the writing of the manuscript, or in the decision to publish the results.

**Figures**

Figure 1

Representative CXR images of COVID-19 pneumonia, non-COVID-19 pneumonia, and the healthy

Abbreviations: chest X-Ray imaging (CXR), novel coronavirus disease (COVID-19)

(A) COVID-19 pneumonia of 30-year-old male

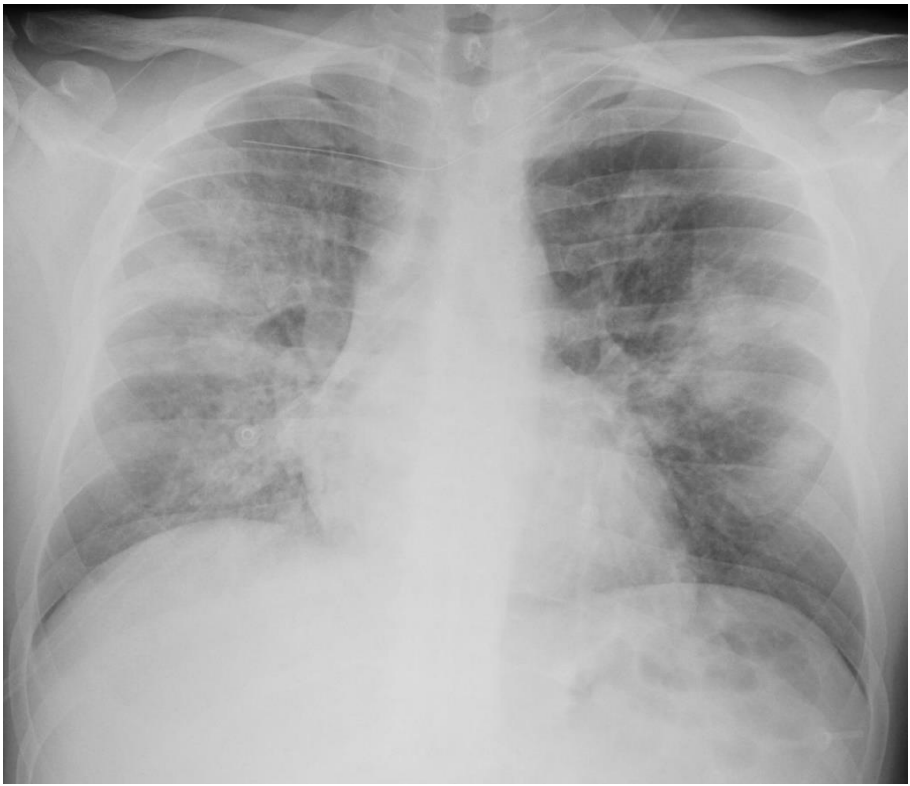

(B) non-COVID-19 pneumonia of 56-year-old male

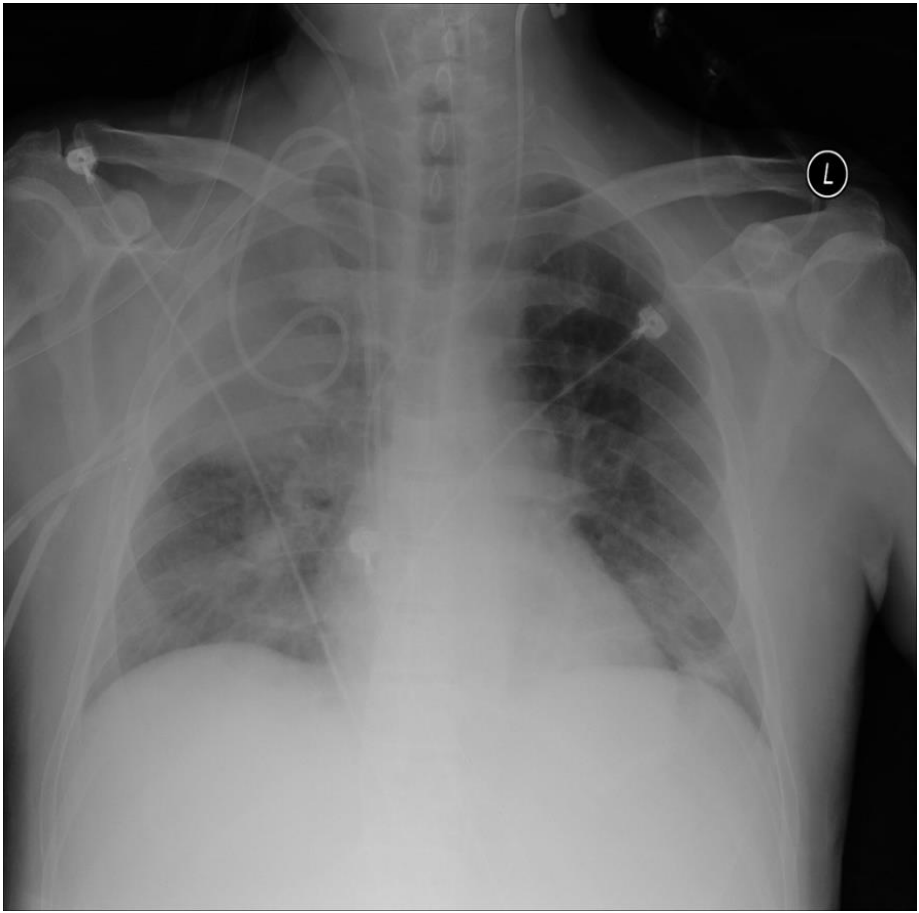

(C) no pneumonia of 60-year-old female

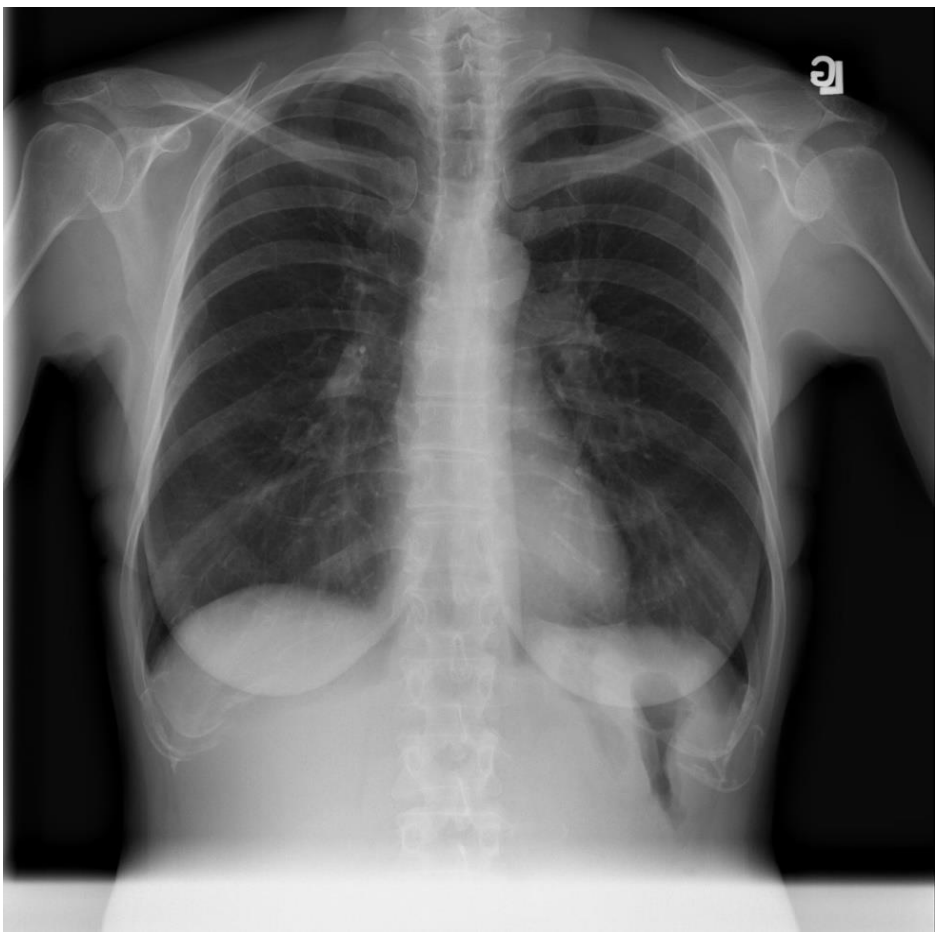

Figure 2

Outline of deep learning model of the proposed method

Note: For pre-trained models, VGG16, Restnet-50, MobileNet, DenseNet-121, and EfficientNet were used in the current study.

Abbreviations: global averaging pooling layer (GAP), fully-connected layer (FC), dropout layer (D)

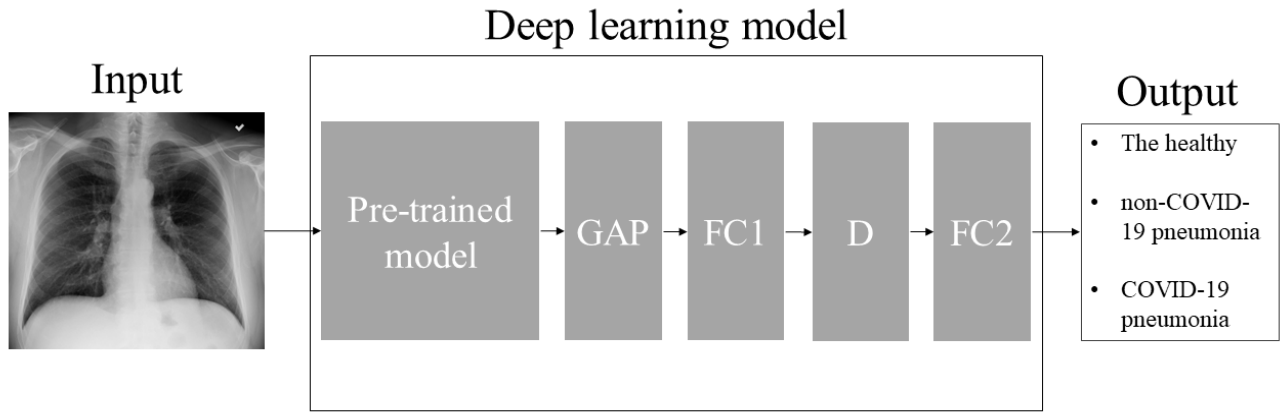

**Tables**

Table 1

Patients' characteristics and CXR attributes.

| Category | | Value |
|---|---|---|
| Number of images | | 1248 |
| Sex | | |
| | Male | 512 |
| | Female | 702 |
| | Not available | 34 |
| Age | | |
| | Available | 1205 |
| | Not available | 43 |
| | Mean ± SD of age (years) | 48.1 ± 17.5 |
| Diagnosis | | |
| | COVID-19 | 215 |
| | non-COVID-19 pneumonia | 533 |
| | the healthy | 500 |
| CXR view | | |
| | PA | 666 |
| | AP | 582 |

Abbreviations: chest X-Ray imaging (CXR), novel coronavirus disease (COVID-19), standard deviation (SD), posterior-anterior view (PA), anterior-posterior view (AP).

Table 2

Results of five pre-trained models

| Models | Loss of test set | 3-category accuracy of test set (%) |
| --- | --- | --- |
| VGG16 (proposed method) | 0.4682 ± 0.0289 | 83.68 ± 2.00 |
| Restnet-50 | 0.5237 ± 0.0161 | 77.76 ± 1.18 |
| MobileNet | 0.4919 ± 0.0300 | 78.72 ± 3.22 |
| DenseNet-121 | 0.5276 ± 0.0082 | 78.24 ± 2.23 |
| EfficientNet | 0.5206 ± 0.0177 | 78.40 ± 1.82 |

Note: Value of each cell was mean ± standard deviation of 5 trials.

Table 3

Representative confusion matrix of 3-category classification in test set

|  |  | Prediction by the proposed model | | |
|---|---|---|---|---|
|  |  | the healthy | non-COVID-19 pneumonia | COVID-19 pneumonia |
| Ground truth | the healthy | 43 | 7 | 0 |
|  | non-COVID-19 pneumonia | 9 | 41 | 3 |
|  | COVID-19 pneumonia | 2 | 0 | 20 |

Note: Accuracy was 83.2% (104/125).

Table 4

Results of ablation study of the proposed method for data augmentation methods and layer freezing

| Models | Loss of test set | 3-category accuracy of test set |
|---|---|---|
| Proposed method | 0.4682 ± 0.0289 | 83.68 ± 2.00 |
| no data augmentation with layer freezing | 0.9009 ± 0.1967 | 78.72 ± 1.65 |
| conventional data augmentation method only with layer freezing | 0.4863 ± 0.0274 | 82.56 ± 2.45 |
| mix-up only with layer freezing | 0.6407 ± 0.0674 | 79.20 ± 1.75 |
| conventional data augmentation method and mixup without layer freezing | 0.5143 ± 0.0179 | 79.04 ± 2.60 |

Note: Value of each cell was mean ± standard deviation of 5 trials.

**Supplementary material**

Table S1

Effect of RICAP as data augmentation method

| Model | Loss of test set |
|---|---|
| proposed method (conventional data augmentation method and mixup with layer freezing) | 0.4682 |
| conventional data augmentation method and RICAP with layer freezing | 0.4724 |
| conventional data augmentation method, mixup, and RICAP with layer freezing | 0.4761 |

Note: Value of each cell was mean of 5 trials.